\documentstyle[twocolumn,aps,prb]{revtex}
\draft
\begin{document}
\begin{title} 
{The critical renormalized coupling constants 
in the symmetric phase of the Ising models}
\end{title}
\author{Jae -Kwon Kim}
\address
{School of Physics, Korea Institute for Advanced Study, \\
207-43 Cheongryangri-dong, Dongdaemun-gu, Seoul 130-012, Korea}
\maketitle

\begin{abstract}
Using a novel finite size scaling Monte Carlo method, we
calculate the four, six and eight point renormalized coupling constants
defined at zero momentum in the symmetric phase
of the three dimensional Ising system. 
The results of the 2D Ising system that were directly 
measured are also reported.
Our values of the six and eight point coupling
constants are significantly different from those obtained
from other methods.
\end{abstract}

\section {INTRODUCTION}
The Hamiltonian of the Ising ferromagnet is given by
\begin{equation}
H = - \sum_{<i,j>} S_i S_j, 
\end{equation}
where the Ising spin at site $i$, $S_i$, can take either 
1 or -1 and the sum is over all the nearest 
neighbors of the lattice.
It is well known that the critical behavior of the D dimensional
Ising model can be described by the D dimensional 
Euclidean scalar field theory the Hamiltonian of which is given by
\begin{equation}
H = \int d^D x \left [ {1 \over 2} (\nabla\phi(x))^2 + {1 \over 2} m_{0}^{2} 
	\phi(x)^2 + {g_{0} \over 4!} \phi(x)^4 \right],
\end{equation}
where $m_0$ and $g_0$ are respectively the bare 
mass and coupling constant defined in the 
absence of critical fluctuations of the fields.  
Near the criticality $m_0$ is
a linear measure of temperature, and we denote 
by $m_{0c}$ the value 
for which the theory becomes critical. 
As the fluctuations become strong 
renormalizations of the mass, coupling
constant and fields are necessary, and the long 
distance behavior is no longer described by 
the bare potential but by 
the effective potential which is generally of more 
complicated functional form than the bare one.
In statistical physics the effective potential represents 
the free energy density as a function of order parameter
(expectation value of the renormalized field), and is 
used to determine the equation of state.

After small-renormalized-field expansion 
of the effective potential, its coefficients are directly 
related to the renormalized coupling
constants (RCCs) defined at zero-momentum. 
In terms of the expectation value of the 
renormalized field, $\varphi_{R}$, 
the effective potential in the 3D symmetric phase
may be written as
\begin{eqnarray}
&&V_{eff}(\varphi_{R}) = {1 \over 2} m_{R}^{2} \varphi_{R}^{2}+ 
     {1 \over 4!}m_{R} g_{R}^{(4)} \varphi_{R}^{4} +\nonumber \\
&&\qquad   {1 \over 6!} g_{R}^{(6)} \varphi_{R}^{6}+
{1 \over 8!} {g_{R}^{(8)} \over m_{R}} \varphi_{R}^{8} + \ldots, 
\end{eqnarray}
where $m_{R}$ and $g_{R}^{(N)}$ represent respectively the 
renormalized mass (inverse correlation length $\xi$) and 
the (dimensionless) $N$-point RCC defined at zero momentum.  

The formal expression of $g_{R}^{(N)}$ can be obtained by
calculating  the $N-$th derivative of the effective potential
with respect to the average value of the renormalized field.
Necessary elements for the calculations are
the well-known relations
\begin{equation}
{dV_{eff} \over d\varphi} = J,~~ 
\varphi = {1 \over V} {dW[J] \over dJ},~~ 
{\rm and}~~\varphi = Z_{\varphi}^{1/2} \varphi_R.
\end{equation}
Here $\varphi$, $V$, $W$, and $Z_{\varphi}$ are respectively
the expectation value of the bare field $\phi$, the volume of the system,
the generating functional for connected Green functions
in the presence of the external field $J$, and the 
field strength renormalization factor given by
$Z_{\varphi} = \chi m_R^{2}$ with 
$\chi$ denoting magnetic susceptibility.
The expressions in 3D are
\begin{eqnarray}
&&g_{R}^{(4)} = -(Z_{\varphi}^{2}/m_{R})~ W_{2}^{-4}~ W_{4}  \\
&&g_{R}^{(6)} = -Z_{\varphi}^{3}~ W_{2}^{-6}~ [W_{6} - 10 
            W_{4}^{2}~ W_{2}^{-1}] \\
&&g_{R}^{(8)}= -m_{R} Z_{\varphi}^{4}~ W_{2}^{-8} \nonumber \\
&& \quad \times [W_{8} - 56W_{6}W_{4}W_{2}^{-1}+280 W_{4}^{3} W_{2}^{-2}],
\end{eqnarray}
where $W_N$ is the Fourier transformed $N$-point connected 
Green function at zero momentum. 
In 2D $\varphi_R$ is dimensionless, so
that the expressions of the $g_{R}^{(2N)}$ 
are given by the above equations 
with additional division of the $m_{R}^2$ term.

Each RCC has a universal zon-zero
finite value as the theory becomes critical, provided
the hyperscaling relation holds. To be more specific,
we take thermodynamic limit first and then take the limit
of the thermodynamic correlation length becoming divergent
(that is, the limit $m_0 \to m_{0c}$).
The corresponding value in this limit may be called as critical RCC.
The determination of the accurate value of 
the critical four-point RCC ($\tilde{g}^{(4)}_R$) 
is particularly important since all other 
universal quantities can be given in terms of it.

In this work we are mainly concerned with
the Monte Carlo calculation of the critical RCCs
in the symmetric phase of the three dimensional 
(3D) Ising model. (From here on the notation of RCC will be
used without the subscript $R$ when it is used in context of
size dependence.) 
This has been a subject of many
studies, including a variety of quantum 
field theoretic approachs\cite{BEN,SOK,GUI,PEL},
high temperature series expansions\cite{BUT,REZ,ZIN}, 
exact renormalization group
flow techniques\cite{BAGN,BERG,MOR}, 
and Monte Carlo simulations\cite{WHE,MAX,PAT}.
It is known that standard direct Monte Carlo
measurements of the thermodynamic critical RCCs
suffer from enormous statistical noise\cite{WHE}.
Other Monte Carlo study is based on the finding of
the probability distribution of the order parameter
in the external field at various temperatures\cite{MAX}. 
We employ a novel finite size scaling (FSS) 
technique\cite{KIM_FSS,LAN} combined with single cluster 
flipping Monte Carlo algorithm\cite{WOL}. 
Our preliminary results
were reported in Ref.(\onlinecite{KIM_LAT96}).
Here we report the result of our updated 
Monte Carlo measurements that has been increased 
more than one magnitude of order. We also report our results
measured at criticality that is useful to determine
the correction to scaling in the RCCs as well as
results in the 2D system.

\section{THE MONTE CARLO METHOD}
For the calculations of the RCC
one needs to calculate $\xi$, $\chi$, and $W_{2N}$.
With periodic boundary condition imposed on the lattice, 
$\chi$ and $W_{2N}$ can be expressed in terms of the
expectation values of various powers of the sum of the spin
over all the lattice site $S\equiv \sum_{i} S_i$.
For example,
\begin{equation}
\chi = <S^2>/L^D,~~W_4=\left(<S^4>-3<S^2>^2\right)/L^D,
\end{equation}
where $L$ is the linear size of the lattice.
$\xi$ can be very accurately determined using the standard
second moment formula\cite{SEE}.
The cluster algorithms\cite{WOL} of the Monte Carlo simulation
have been extremely efficient for many problems of 
critical phenomena. It is now an easy task to obtain
Monte Carlo data of typical physical quantities such as $\chi$ and $\xi$
with relative statistical errors less than 0.1 percent
at a temperature arbitrarily close to criticality.
Nevertheless, it turns out that accurate Monte Carlo measurement
of higher order RCC is problematic 
in some cases. The problem stems basically from the fact that
the higher order RCC is given as a multiplication
of a huge number with a tiny one.
The former is given typically by some power of
${L \over \xi_L}$, for instance $\left(L \over \xi_L\right)^{9}$ 
for the $g_{R}^{(8)}$ of the 3D system, and the latter
comes from the combination of the $W_{2N}$
that turns out to be extremely sensitive to 
statistical noises. The noises increase rapidly with
increasing temperature and increasing value of $L/\xi_L$.

To illustrate we measured various physical quantities 
at an arbitrary (inverse) temperature in the scaling regime of the 2D
Ising model with increasing linear size of the lattice L.
Table(\ref{tab:2D_420}) clearly shows that the statistical 
error in the measured values of $g_{R}^{(2N)}$ increases with $N$ 
for any given value of L.
For L=80, for example, the relative statistical errors
of $\xi_L$, $\chi_L$, $g^{(4)}_{L}$, $g^{(6)}_{L}$, $g^{(8)}_{L}$
are respectively 0.08, 0.1, 0.7, 3.1, 5.3 percents.
We observe clearly that all the variables except for $g^{(8)}_L$
are monotonically increasing functions of $L$,  up to
L=80 where $L/\xi_L \simeq 6.7$. 
For further increasing $L$ the values do not vary
within the statistical errors. 
We recall that the well-known exact thermodynamic value of 
the $\xi$ at this $\beta$ is 11.9055..., which manifests itself
at L=80 already within the statistical errors.
The size independent value thus corresponds to the thermodynamic value.
It is also very clear that the size dependence becomes rapidly weaker
with increasing $L$ for all the variables we consider here.
This is expected to be the case for any physical variable having
well-defined thermodynamic value.
For the $g^{(8)}_{L}$, however, due to the large statistical noises
is it hard to draw a definite conclusion on the size dependence.
Nevertheless, it is very natural to expect that
$g^{(8)}_L$ would follow similar size dependence as
$g^{(4)}_L$ and $g^{(6)}_L$.
 
To compute the critical RCC it is required to obtain thermodynamic
values as $\beta \to \beta_c$. For the case of the 2D Ising model,
we will not follow that procedure. Instead, we will report
the results of measurements at criticality and try to find out
possible effect of the correction to scaling. 
The illustration in the case of
the 2D Ising model will be useful in connection 
with the study of the 3D case.
Note that the relation in the scaling regime
\begin{equation}
g^{(2N)}_{R}(t) \sim t^{D\nu-2\Delta_{2N}+\gamma},
\end{equation}
translates into
\begin{equation}
g^{(2N)}_{L}(t=0) \sim L^{(D\nu-2\Delta_{2N}+\gamma)/\nu},
\end{equation}
at criticality. Hence the hyperscaling relation 
$D\nu-2\Delta_{2N}+\gamma=0$
implies the invariance of $g^{(2N)}_{L}(t=0)$ with respect to $L$.
Conversely, from the (weak) dependence of of $g^{(2N)}_{L}(t=0)$
can one infer the correction to scaling in the scaling regime.
Computing RCC by taking the limit $t\to 0$ first and then 
take the limit $L \to \infty$ is equivalent to the computation
of $g^{(2N)}_{L}(t=0)$. It should be stressed that $g^{(2N)}_{L}(t=0)$ 
basically represents the apposite of thermodynamic limit for a non-zero
$t \to 0$ (that is $L \to 0$ limit for $\xi$ becomes huge but not infinite)
and has nothing to do with the 
lower-bound of critical RCC\cite{KIM_COM}.

We also report so called 6-th and 8-th order cumulant ratios 
at criticality that are denoted by $U^{(6)}_L$ and $U^{(8)}_L$ respectively.
These are part of $W_{6}$ and $W_{8}$, and 
have already been available for the 2D Ising medel
based on umbrella sampling method of 
the Monte Carlo simulation\cite{MON}. 
Since our code is generic in dimension,
the comparison with the result from different
Monte Carlo method for these higher order
cumulant ratios having severe statistical noises may be useful
to check against any unexpected possible systematic errors
in our Monte Carlo measurements.
For example, $U^{(8)}_L$ is expressed by
\begin{eqnarray}
U^{(8)}_L = &&(<S^8>-28<S^6><S^2>-35<S^4>^2+\nonumber \\
            && 420<S^4><S^2>^2-630<S^2>^4)/<S^2>^4
\end{eqnarray}
The results are summerized in Table(\ref{tab:2D_FSS}).
Our results of $U^{(2N)}_L$ for N=2,3, and 4 are
respectively 1.8318(5), 13.93(1), and -226.0(1), which may be
compared with 1.834, 13.96, and -226.6 reported 
in Ref.\onlinecite{MON}. 
Currently accepted value of $U^{(4)}_L$ is 1.832(1). 
We observe that $g^{(2N)}_L(t=0)$ has no $L$ dependence at least for
$L \ge 40$, confirming the hyperscaling and 
no significant correction to scaling. We therefore lead to 
the conclusion that the thermodynamic values of RCC measured at
$\beta=0.420$ may well be regarded as the critical values.

Regarding size dependence 
we observe similar feature in the 3D Ising model
as in the 2D case. Our data are summerized in
Table(\ref{tab:3D_217}). For the $\beta=0.217$ and 0.220 
we observe no size dependence (within the statistical errors) for
the $\xi$ and $\chi$ beyond L=32 and 60 respectively.
This roughly corresponds to $L/\xi_L \simeq 5.5$.
We expect that this is the case even for the RCCs, although
it was almost impossible to get precise measurements
of the thermodynamic values of $g^{(6)}_L$ and $g^{(8)}_L$
for this value of $L/\xi_L$. 
For example, for $\beta=0.217$ 
and $L=32$ we generated about $10^9$ single cluster sweeps,
but error bars are larger than the mean value for the $g^{(8)}_L$.
The values of $g^{(6)}_L$ for $L \ge 32$ also seem to be 
unreliable in view of the generic feature of the weaker 
size dependence with larger $L$.

It thus appears that Monte Carlo computation of the 
critical RCCs relying on direct brute force measurements
is prohibitively difficult. 
In order to overcome the difficulty closer to $T_{c}$,
we make use of a new FSS
function ${\cal{Q}}_{A}(x(L,t))$\cite{KIM_FSS,LAN},
defined by the expression
\begin{equation}
A_{L}(t)=A(t){\cal{Q}}_{A}(x(L,t)),~~x(L,t)\equiv \xi_{L}(t)/L.  
\label{eq:fun}
\end{equation}
Here $A_{L}(t)$ represents the quantity $A$ measured on a finite
lattice of linear size $L$ at a reduced temperature $t$,
with its corresponding thermodynamic value $A(t)$.
What Eq.(\ref{eq:fun}) states is that
the size dependence of a physical quantity $A$ is given 
as a function of the scaling variable $x$. 
As a result, the ratio of $L$ to $\xi_L$ beyond which
thermodynamic limit is reached is independent of the temperature,
that turned out to be approximately 5.5 for the 3D Ising model.

The FSS technique is especially useful for our purpose, because
it enables us to extract accurate thermodynamic values based
on the Monte Carlo measurements with much smaller lattices. 
We just outline the single step FSS extrapolation technique used in this work. 
For a detailed explanation, 
we refer the readers to Ref.\onlinecite{KIM_FSS,LAN}.
\begin{enumerate}
\item For a certain $t_{0}$, measure $A_{L}(t_{0})$ and 
       $x(L,t_{0}) = \xi_{L}(t_{0}) /L $ for increasing L.  
\item Determine the thermodynamic value
      at the temperature $A(t_{0})$ by measuring $A_{L}(t_0)$ 
      which is $L$ independent. 
\item Fit $(x(L,t_{0}), A_{L}(t_{0})/A(t_{0}))$ data to a
       suitable functional form. 
       In this work we used the ansatz,
	\begin{equation}
       {\cal Q}(x) = 1+c_{1} x+c_{2}x^2+c_3x^3+c_4 x^4 \label{eq:ans}
	\end{equation}
\item For any other $t$, choose a suitable L, measure the value of 
       $x(L,t) \equiv \xi_{L}/L$ and $A_{L}(t)$,
       and interpolate ${\cal Q}(x(L,t))$.  
\item Extract $A(t)$ by plugging $A_{L}(t)$ and ${\cal Q}(x(L,t))$
       into Eq.(\ref{eq:fun}).  
\end{enumerate}
The smallest value of L we considered for the FSS method is 20.

\section {RESULT AND DISCUSSION}
Our choice of $\beta_0$ is $\beta_0=0.220$.
We infer from the generic feature of the size dependence observed for 
the 3D Ising model that thermodynamic limit of the RCCs are reached
for $L/\xi_L \gtrsim 5.5$. We thus get $g^{(4)}_R = 24.5(4)$
and $g^{(6)}_R = 1983(148)$ at this $\beta$. 
Fitting the data to the ansatz Eq.(\ref{eq:ans}), 
we get
\begin{eqnarray*}
&&c_1=2.338,~~c_2=-15.768,~~c_3=19.770,~~c_4=-5.123 \\
&&c_1=3.225,~~c_2=-25.046,~~c_3=42.485,~~c_4=-23.177
\end{eqnarray*}
for the scaling function ${\cal Q}(x)$ of $g^{(4)}$ and $g^{(6)}$
respectively. 
Using the scaling function we calculated
the thermodynamic values of the 4-point and 6-point RCCs
for all the L from 36 to 80 in the Table(\ref{tab:3D_221}).
The result at each L is in reasonably good agreement. For example,
we get $g^{(4)}_R$= 24.3(1), 24.4(4), 24.1(1), 23.9(1),
23.9(1), 24.1(3), and 23.9(2) for each $L$ 
from the L=36 through the L=80 in the table.
For $g^{(6)}_R$ we get 1919(11), 1939(20), 1939(21), 1915(10),
1906(17), 1917(70), and 1897(49). The invariance of the thermodynamic
RCC with respect to the choice of $L$ is a numerical proof
of the FSS for the variables (see Figure(\ref{fig:g4}) and (\ref{fig:g6})). 
We usually extracted the thermodynamic value for several different
choices of $L$ for a given temperature, and took the average.
Our net results from $\beta=0.217$ to $\beta=0.2213$ are found
in Table(\ref{tab:net}). It is observed that both $g^{(4)}_R$ and
$g^{(6)}_R$ tend to decrease mildly as $\beta \to \beta_c$.
In this work we assume the widely accepted correction to scaling exponent 
$\theta \simeq 0.5$ and $\beta_c=0.221654$\cite{ZINN}. 
By fitting our data in Table(\ref{tab:net}) to
\begin{equation}
g^{(2N)}_R(t) = \tilde{g}^{(2N)}_R (1 + a_{2N} t ^{0.5}), \label{eq:theta}
\end{equation}
we obtain the critical RCC which reads
\begin{eqnarray}
&&\tilde{g}^{(4)}_R = 23.6(2) \\
&&\tilde{g}^{(6)}_R = 1879(50)
\end{eqnarray}

Our results of $g^{(2N)}_L(t=0)$ are found in Table(\ref{tab:3D_FSS}).
It is observed that both of $U_L^{(4)}$ and $\xi_L/L$ have tendency
of very mild decreasing with increasing $L$. 
All the $g^{(2N)}_L(t=0)$ show remarkable invariance with respect to
increasing L at least for $L \ge 30$. In other words, they do not
show the effect of correction to scaling observed in the scaling regime.
There may be a few possible interpretations for the discrepancy.
First let us remind that Eq.(\ref{eq:theta}) translates into
\begin{equation}
g^{(2N)}_L(t=0) = \tilde{g}^{(2N)}(t=0) (1 + b_{2N} L ^{-\omega}),
~~(\omega > 0),
\end{equation}
and that the coefficient $b_{2N}$ may happen to be very small.
The second possibility is that currently accepted $\beta_c$
might be slightly underestimated: At exact criticality it is expected
that $\xi_L / L$ remains an exact constant irrespective of the
presence of the correction, which seems not to be 
the case at $\beta=0.221654$.

Our current result at criticality may be summerized as follows.
\begin{eqnarray}
&&\tilde{g}^{(4)}(t=0) = 5.36(2)   \\
&&\tilde{g}^{(6)}(t=0) = 154.6(1.1) \\
&&\tilde{g}^{(8)}(t=0) = 1.035(10) \times 10^{4}.
\end{eqnarray}

It is impossible to apply the FSS
method for the $g^{(8)}_R$ due to its large error bars.
Nevertheless, it is highly likely that at least $\tilde{g}^{(8)}_R$ is 
of order $10^5$. (see Table(\ref{tab:3D_217}).)
Also we have reliable data  at $\beta=0.2213$ showing
$g^{(8)}_L \simeq  1.13(3) \times 10^5$ already for $L/\xi_L \simeq 2.7$.
Furthermore it is observed that the ratio of critical RCCs to
$g^{(2N)}_L(t=0)$ tends to increase with $N$; 
The ratio is roughly 4.4 and 12.2 for N=2 and 3 respectively for the 3D. 
We also note that $L$ dependence of $g^{(8)}_L(t=0)$  
is quanitatively the same as that of the other RCCs, indicating that
the correction to scaling for $g^{(8)}_R(t)$ is as mild as the other RCCs.
All the evidence almost undoubtedly point out that
the critical $g^{(8)}_R$ is much larger than those estimated 
by other methods. 
Our crude estimate of $\tilde{g}^{(8)}_R$ reads
\begin{equation}
\tilde{g}^{(8)}_R \simeq 1.4(3) \times 10^5.
\end{equation}

Our value of $\tilde{g}_{R}^{(4)}$ is in reasonable agreement with
other estimates.  The agreement is especially good with the result
from high-temperature expansion\cite{BUT} and the field theoretic
treatment\cite{GUI}. However, the agreement 
becomes worse as $N$ increases: The results of $\tilde{g}^{(6)}_R$
from previous studies are within the range 
$860 \lesssim \tilde{g}^{(6)}_R \lesssim 1515$. The closest result to ours, 
$\tilde{g}^{(6)}_R \simeq 1515$, obtained from the previous Monte Carlo
method\cite{MAX} is remarkable because it is obtained not by
the measurement of RCC but by the  Monte Carlo
measurement of the probability distribution of order parameter assuming
the thermodynamic limit for $L/\xi_L \simeq 4$. We conjecture that this result
would agree much better with our estimate if the same thermodynamic limit as
in this work would have been taken. 
The values of $\tilde{g}^{(8)}_R$ from previous studies ranges 
from $2.9 \times 10^4$ to $3.5 \times 10^4$, 
which is at least four times smaller than our estimate.
Although it was pointed out\cite{BUT} that longer series terms are 
necessary for  more accurate estimate of higher order critical RCC from the
high temperature series expansion method, it remains mysterious
why previous studies based on different methods give rise to reasonably 
consistent results among them. Nevertheless, in view of limitations
of almost all the methods used to study this subject, it is fair to say that
the issue is still open to further studies.

Our results of the critical RCCs for 2D Ising model obtained 
assuming negligibly small correction to scaling read
\begin{eqnarray}
\tilde{g}^{(4)}_R &=& 14.7(2) \\
\tilde{g}^{(6)}_R &=& 850(25) \\
\tilde{g}^{(8)}_R &=& 8.9(5) \times 10^4.
\end{eqnarray}
The only result on the high order RCCs 
we  are aware of  is the $\varepsilon$ expansion study\cite{VICA};
\begin{equation}
\tilde{g}^{(6)}_R =  794,~~~\tilde{g}^{(8)}_R \simeq 8.2(2) \times 10^4,
\end{equation}
which agree reasonably well with ours.

The results at criticality are summerized as
\begin{eqnarray}
\tilde{g}^{(4)}(t=0) &=& 2.239(7) \\
\tilde{g}^{(6)}(t=0) &=& 29.34(20) \\
\tilde{g}^{(8)}(t=0) &=& 947(10) 
\end{eqnarray}

We wish to thank Maxim Tsypin for many communications.

\begin{table}[t]
\caption{Size dependence of various physical quantities
         at $\beta=0.420$ up to L=100 for the 2D
         Ising system.}
\label{tab:2D_420}
\begin{tabular}{llllll}
L  &$\xi_L$   &$\chi_L$   &$g^{(4)}_{L}$ &$g^{(6)}_{L}$ &$g^{(8)}_{L}$  
\\ \hline
20   &9.664(8)   &116.1(1)    &6.29(1)   &218.46(1.32)    &17871(140) \\
30   &10.977(9)  &162.5(1)    &8.95(1)   &414.8 (2.3)     &43532(313) \\
40   &11.53(1)   &186.5(2)    &11.26(1)  &615.1(4.0)      &71443(615) \\
50   &11.78(1)   &197.4(2)    &12.83(3)  &740.2(5.5)      &85462(879)\\
60   &11.84(1)   &201.3(2)    &13.79(4)  &794.8(6.2)      &87919(1185) \\
70   &11.88(1)   &203.1(2)    &14.20(6)  &825.1(10.0)     &89824(2315) \\
80   &11.91(1)   &203.9(2)    &14.77(10) &850.4(26.1)     &89521(4765) \\
90   &11.91(2)   &204.3(2)    &14.85(12) &858.8(24.2)     &90722(7451) \\
100  &11.90(2)   &204.4(3)    &14.60(16) &846(38)         &88165(9575) 
\end{tabular}
\end{table}

\begin{table}
\caption{Size dependence of various physical quantities
         at criticality $\beta_c=\ln~(\sqrt{2}+1)/2$ up to L=100 for the 2D
         Ising system. }
\label{tab:2D_FSS}
\begin{tabular}{lllllll}
L  &$U^{(4)}_L$   &$U^{(6)}_L$   &$U^{(8)}_L$   &$g^{(4)}_{L}$
                       &$g^{(6)}_{L}$ &$g^{(8)}_{L}$ \\ \hline
20  &1.8324(6)  &13.94(1)  &-226.1(2)   &2.227(7) &29.03(18)  
                                        &933(9)  \\
40  &1.8321(6)  &13.94(1)  &-226.1(2)   &2.237(8) &29.28(21)
                                         &945(10)  \\
60  &1.8317(5)  &13.93(1)  &-225.9(1)   &2.241(7) &29.37(18)
					&949(9)  \\
80  &1.8318(5)  &13.93(1)  &-226.0(1)  &2.240(6) &29.36(16)
                                        &948(8)  \\
100 &1.8316(6) &13.93(1)  &-226.0(1)  &2.239(7) &29.33(18)
				      &947(9) 
\end{tabular}
\end{table}

\begin{table}
\caption{Size dependence of the various physical quantities at
         $\beta=0.217$ (the upper part) and $\beta=0.220$ (the lower part).
	 Note that the $g^{(8)}_{L}$ data for $L/\xi_{L} \ge 5.7$ become
         unreliable due to huge error bars and in view of weaker size dependence
	 for larger $L$. This appears to be case even to the $g^{(6)}_{L}$
         for $L=36$. }
\label{tab:3D_217}
\begin{tabular}{lllllc} 
L    &$\xi_{L}$  &$\chi_{L}$ &$g^{(4)}_{L}$ &$g^{(6)}_{L}$  
&$g^{(8)}_{L}\times 10^{-4}$ \\ \hline
8   &3.93(0)  &59.49(6)   &9.86(2)  &492.5(1.9) &5.45(3) \\
12  &4.85(1)  &94.16(12)  &13.59(3)  &852(4)    &10.8(1)\\
16  &5.299(1) &114.54(5)  &17.71(2)  &1277(3)  &15.77(9)\\
20  &5.488(3) &124.20(7)  &21.07(5) &1562(9)  &15.5(3)\\
24  &5.573(2) &128.40(5)  &23.16(8) &1735(22)  &15.5(9)\\
28  &5.605(4) &130.06(11) &24.72(11) &1887(53)  &14.6(3.8)\\
32  &5.622(2) &130.87(5)  &24.88(14) &2190(86) &13.3(16.9) \\
36  &5.619(19)&130.96(14) &25.64(16) &2860(191) &83.3(29.1) \\ \hline
16  &7.85(2)  &228.8(7)	  &9.52(5)   &452.1(4.6) & -  \\
20  &8.85(2)  &298.1(7)    &11.43(5)  &625.6(5.1) & -   \\
24  &9.56(2)  &351.8(1.1)  &13.44(7)  &814.7(8.2)  & -  \\
30  &10.20(3) &407.1(1.1) &16.5(1)   &1123(16)    & -  \\
36  &10.56(2) &439.2(1.2)   & 19.2(1)  &1372(27) & -  \\
40  &10.68(3) &455.2(1.5)   & 21.4(2)  &1501(56)  & - \\
50  &10.83(3) &467.9(1.3)   & 23.7(3)  &1795(109)  & - \\
60  &10.89(3) &472.3(1.8)   & 24.5(4)  &1983(148)  & - \\
70  &10.90(3) &473.0(1.1)   & 25.8(1.6)  & -        & - 	
\end{tabular}
\end{table}

\begin{table}
\caption{Size dependence of the various physical quantities
         at $\beta=0.221$ up to $L/\xi_{L} \simeq 4.11$.}
\label{tab:3D_221}
\begin{tabular}{lllllc} 
L &$\xi_{L}$ &$\chi_L$ &$g^{(4)}_{L}$   &$g^{(6)}_{L}$ 
&$g_{L}^{(8)}\times 10^{-4}$ 
\\ \hline 
20 & 10.95(2)  &426.7(1.4) &7.6(1)  &301(3)   &2.67(4)  \\
28 & 13.93(6)  &703.4(4.0) &9.2(1)  &423(8)   &4.3(1)  \\
36 & 15.98(6)  &944.2(4.7) &11.2(1) &600(10)  &6.6(2)   \\
40 & 16.74(5)  &1045.5(4.3) &12.4(1) &710(13) &8.2(2)   \\
48 & 17.91(3)  &1210(3)    &14.59(6) &920(8)  &10.8(2)  \\
56 & 18.60(1)  &1316(1)    &16.74(5) &1116(7) &12.2(2)  \\
64 & 19.03(2)  &1386(2)    &18.66(7) &1297(13) &13.4(4) \\
72 & 19.28(5)  &1426(6)    &20.47(29) &1465(57) &13.3(2.6) \\ 
80 & 19.46(2)  &1458(2)    &21.36(16)&1580(82) & -
\end{tabular}
\end{table}

\begin{table}
\caption{Size dependence of various physical quantities
         at criticality $\beta_c=0.221654$ up to L=80 for the 3D
	 Ising system. Here $U^{(4)}_L$ represents fourth-order 
         cumulant ratio. }
\label{tab:3D_FSS}
\begin{tabular}{lllllll}
L  &$U^{(4)}_L$   &$\xi_L$   &$\chi_L$   &$g^{(4)}_{L}$ &$g^{(6)}_{L}$
                           &$g^{(8)}_{L} \times 10^{-4}$  \\ \hline
20 &1.418(1) &12.80(1) &545.1(5)    &5.40(1) &157.4(8)    &1.07(1) \\
30 &1.409(1) &19.20(2) &1215(1)     &5.37(1) &155.3(7)    &1.04(1) \\
40 &1.408(2) &25.65(4) &2148(4)     &5.34(2) &153.5(1.1)  &1.03(1) \\
50 &1.403(1) &31.98(4) &3313(4)     &5.36(2) &154.4(1.0)  &1.03(1) \\
60 &1.401(2) &38.33(6) &4742(10)    &5.37(2) &155.1(1.1)  &1.04(1) \\
70 &1.399(3) &44.69(8) &6418(19)    &5.38(2) &155.2(1.2)  &1.04(1) \\
80 &1.398(2) &51.13(9) &8351(19)    &5.36(2) &154.1(1.3)  &1.03(1) 
\end{tabular}
\end{table}

\onecolumn
\begin{table}
\caption{Thermodynamic values of the four and six point
         RCC extracted by the 
         FSS technique for some temperatures over
         $0.217 \le \beta \le 0.2213$. Here the quoted errors
         are obtained ignoring the statistical errors in the estimate
         of the thermodynamic values at $\beta=0.220$. 
         Crudely speaking,
         increases in the estimate of $beta=0.220$ would lead to
         more or less similar amount of increase in the estimate for
	 other temperatures.  }
\label{tab:net}
\begin{tabular} {l|cccccr}
$\beta$  &0.217   &0.219  &0.220  &0.2206  &0.2210   &0.2213 \\  \hline
$g_{R}^{(4)}$ &25.2(2) &24.7(2) &24.5(0) &24.2(2) &24.2(2) &24.1(2)\\
$g_{R}^{(6)}$ &2034(69) &2063(46) &1983(0) &1988(81) &1919(51) &1943(42) \\
\end{tabular}
\end{table}
\twocolumn

\begin{figure}
\caption{${\cal Q}_{g^4}$ calculated for $\beta$=0.220 and 0.221. The two data set collapse
		unto a single universal curve, showing a numerical evidence for
		the FSS \protect{Eq.\ref{eq:fun}} for $g_{R}^{(4)}$. }
\label{fig:g4}
\end{figure}
\begin{figure}
\caption{ The same as in Figure(\ref{fig:g4}) but for ${\cal Q}_{g^6}$.}
\label{fig:g6}
\end{figure}

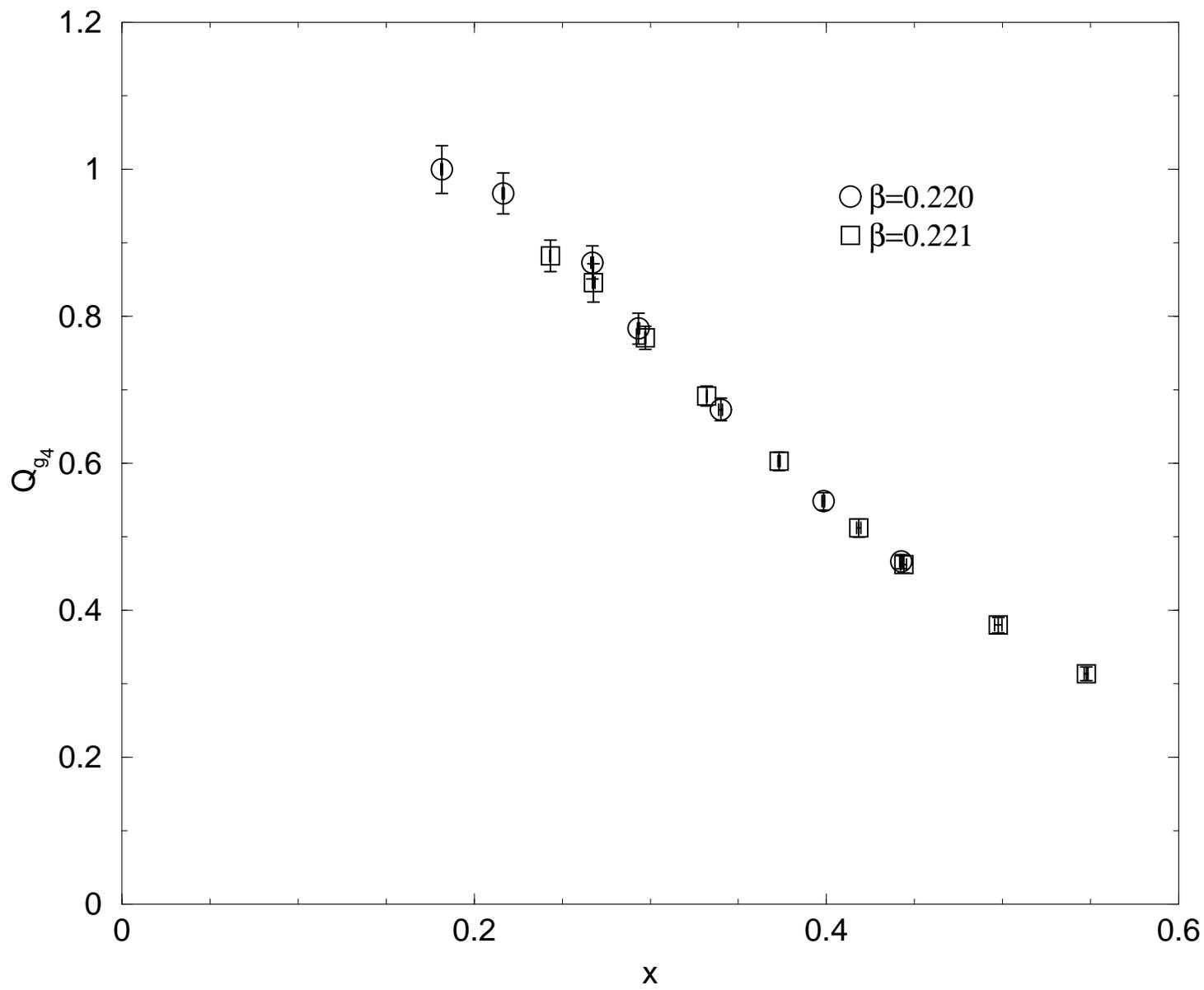
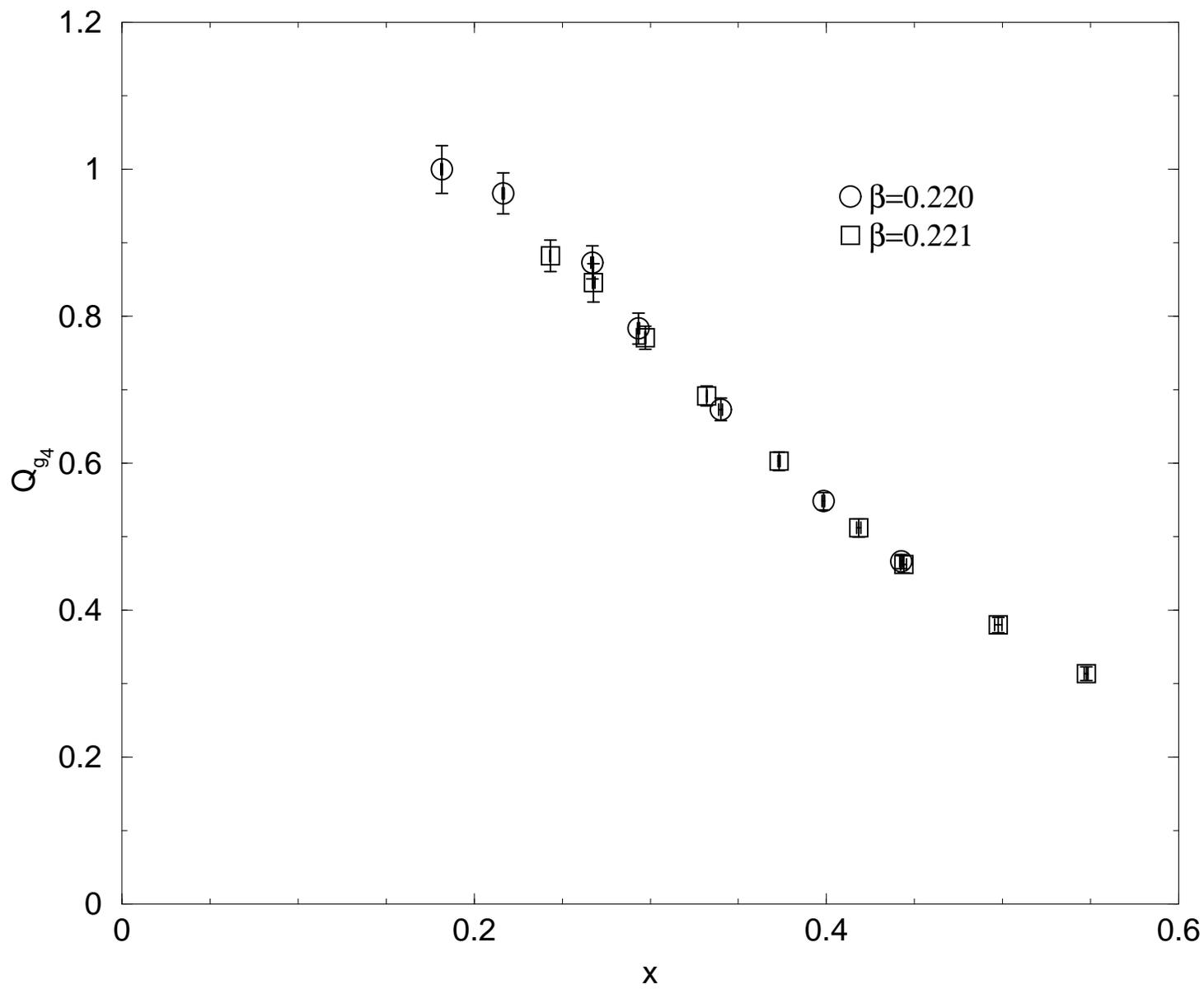
\end{document}